\documentclass[12pt,a4paper]{article}

\makeatletter
\@addtoreset{equation}{section}
\makeatother

\begin{document}

\baselineskip16pt

\begin{flushright}
\footnotesize
\footnotesize
CERN-TH/2000-374\\
{\tt hep-th/0012137}\\
December, $2000$
\normalsize
\end{flushright}

\vspace{8mm}

\begin{center}

\vspace{.8cm}
{\Large \bf Non-commutative Branes from M-theory} 

\vspace{1cm}

Yolanda Lozano

\vspace{.5cm}

{
{\small \sl Theory Division, CERN\\
1211 Gen\`eve 23, Switzerland}\\
{\tt yolanda.lozano@cern.ch}
}

\vspace{.4cm}

\vspace{2cm}


{\bf Abstract}

\end{center}
\begin{quotation}

\small

The analysis of the worldvolume effective actions of the M-theory
Kaluza-Klein monopole and 9-brane suggests that
it should be possible to describe non-abelian configurations of M2-branes
or M5-branes if the M2-branes are transverse to the
eleventh direction and the M5-branes are wrapped 
on it. This is determined by the fact that the Kaluza-Klein
monopole and the M9-brane are constrained to move in particular isometric
spacetimes. We show that the
same kind of situation is implied by the analysis of the brane
descent relations in M-theory.
We compute some of the non-commutative couplings of the
worldvolume effective actions of these non-abelian systems 
of M2 and M5 branes and show that they
indicate the existence of configurations corresponding to N branes expanding
into a higher dimensional M-brane. The reduction to Type II brings up
new descriptions of coincident D-branes at strong coupling.
We show that these systems have the right non-commutative charges
to describe certain expanded configurations playing a role in the
framework of the AdS/CFT correspondence. Finally, we discuss the
realization of non-commutative brane configurations as topological
solitons in non-abelian brane-antibrane systems.\\
\\

\end{quotation}


\newpage

\section{Introduction and Summary}

Non-abelian D-brane systems have recently attracted a lot of attention. 
In remarkable papers Myers and Taylor and Van Raamsdonk
observed that a system of N coincident D$p$-branes can develop 
multipole moments under 
Ramond-Ramond fields that would normally couple to higher dimensional branes
\cite{Myers} \cite{TvR}.
This is possible because the non-abelian D$p$-brane embedding 
coordinates become non-commutative. An external field $C^{(q+1)}$, $q>p$,
($p$, $q$ even (odd) in IIA (IIB))
can polarize the $p$-branes to expand into a non-commutative 
configuration that can be interpreted as a D$q$, N D$p$ bound state.
Alternatively the original D$p$-branes can be represented as a 
single D$q$-brane with N units of 
worldvolume instanton-like density \cite{Douglas}.
Both approaches agree in the large N limit \cite{Myers,CMT}.

The `dielectric' property has been shown to play an important role 
within the AdS/CFT correspondence. Dielectric branes have been used
to find non-singular string theory duals of gauge theories living on
D-branes with reduced supersymmetry \cite{PS,AR,Bena}.
In \cite{PS} the supergravity dual of a four dimensional 
${\cal N}=1$ confining gauge theory, obtained by perturbing the
${\cal N}=4$ gauge theory living on N D3-branes, was identified
as a non-singular spacetime with an expanded brane source 
arising from Myers' 
dielectric effect. In particular, a mapping between the
gauge theory vacua and states corresponding to D3-branes being
polarized into D5 and NS5 branes, with worldvolume $R^4\times S^2$,
in $AdS_5\times S^5$ was found, with the Higgs vacuum
represented by a single D5-brane configuration and the confining
vacuum by an NS5-brane.
Similar issues have also been investigated in an M-theory 
framework. A perturbation of the ${\cal N}=8$ three dimensional 
gauge theory living on N M2-branes to ${\cal N}=2$
has been shown to be dual to
M2-branes expanding into an M5-brane of geometry $R^3\times S^3$
in $AdS_4\times S^7$ \cite{Bena2}.
The existence of polarized M2-branes is expected from duality. 
However, the system cannot be described as a collection of branes 
in an external field developping a dipole moment and expanding,
given that the degrees of freedom of the worldvolume theory of
coincident M2-branes are not
known. The approach of reference \cite{Bena2} was to consider the 
alternative description as an 
M5-brane in $AdS_4\times S^7$
with a non-trivial flux on $S^3$, carrying 
M2-brane charge N. This was also the approach of \cite{Bena}, where
the dual of the confining vacua of a perturbed three dimensional
gauge theory living on D2-branes was identified as
D2-branes polarized into an NS5-brane. 

Expanding N M0, M$(n-2)$ systems have also been proposed to describe 
gravitons carrying angular momentum in $AdS_m\times S^n$,
with the M$(n-2)$-brane moving on the $S^n$ sphere \cite{GST}. 
The `giant' gravitons of \cite{GST} in $AdS_7 \times S^4$ are 
identified with (N M0, M2)
bound states in a constant four-form magnetic field strength,
where the M2-brane expands on the $S^4$ of $AdS_7\times S^4$
\cite{DTV}. Similarly (N M0, M2) bound states in an electric four-form
field strength can be constructed such that the M2-branes expand into 
the $AdS_4$
component of $AdS_4\times S^7$. Configurations of (N M0, M$(m-2)$) branes
in which the M$(m-2)$-brane expands out into
the $AdS_m$ component of the spacetime have been associated to
`dual giant gravitons' (see \cite{GMT}).

In this paper we show that it is possible to understand all
these configurations in terms of non-abelian branes 
expanding into a higher dimensional brane. We start in section 2 by
summarizing the main results in \cite{Myers,CMT}, putting 
special emphasis in the interpretation of the couplings 
in the Wess-Zumino action describing the non-abelian system
in terms of expanding configurations.
In section 3 we study non-abelian
brane configurations in M-theory. By analyzing the couplings present
in the Kaluza-Klein monopole and M9-brane effective actions we give
an indication of the special type of
non-abelian M2 and M5 configurations that can arise as topological 
solutions in M-theory. The M2-branes appear delocalized in the 
eleventh direction, whereas the M5-branes are wrapped on it. 
This situation is consistent with the fact that in the non-abelian
case the M2 and M5 branes should be able to expand into the Kaluza-Klein
monopole and the M9-brane, which see the eleventh direction
as a special isometric direction. The configurations of M2 and M5
branes that we find were shown to
play a role as supergravity duals of ${\rm (S)YM}_{3,6}$ 
in certain regions
of the parameter space \cite{IMSY}.

We will propose a worldvolume Wess-Zumino effective action describing 
non-abelian M2 and M5 branes.
The requirement is that they should reproduce
the effective actions of non-abelian D2 and D4 branes upon
reduction to Type IIA along the special isometric direction. 
We will interpret the terms that couple in the effective action as
associated to non-commutative configurations corresponding to the
branes expanding into higher dimensional branes. Among these
we find the case in which N M2-branes expand into an
M5-brane, considered in \cite{Bena2}. We will also discuss the
worldvolume effective action describing coincident M0-branes, and
we will identify the couplings responsible for the configurations
in which N M0-branes expand
into M2 and M5 branes \cite{GST,GMT,DTV}. 

In section 4 we show that the non-abelian configurations that we have
constructed in M-theory give rise, upon reduction to Type IIA, 
to strongly coupled D$p$-brane systems with the right
worldvolume couplings to
explain some non-abelian configurations that have been identified
as supergravity duals of certain gauge theories living on D-branes.
We will see for instance that a delocalized D2-brane system contains
a coupling $i_\Phi i_\Phi i_k B^{(6)}$ \footnote{$\Phi$ are the
embedding scalars and $B^{(6)}$ the NS-NS 6-form. See the notation
in the next section.} in its effective action, describing the
N D2, NS5 bound state considered in \cite{Bena}. This system is
related by T-duality with the N D3, NS5 bound state of \cite{PS}.
In section 5 we show that the strongly coupled D$p$-brane systems
constructed in Type IIA by reduction from M-theory are connected to
strongly coupled D$p$-brane systems in Type IIB. Delocalized N D2-branes
are mapped for instance onto N D3-branes at strong coupling. The reason a 
distinction must be made between weakly coupled and strongly coupled
D3-branes is that in the non-abelian case the corresponding actions
cannot be shown to be related by an SL(2,Z) worldvolume duality
transformation. This has the implication that non-abelian D2-branes at
strong coupling should only have manifest O(6) transverse rotational 
symmetry, which is consistent with the fact that the strongly coupled
D2-branes that we can construct by reduction from M-theory are
delocalized in one of the space directions, derived from the fact that
our M2-brane system is O(7) $\subset$ O(8) (transverse) rotationally
invariant.
There are arguments however showing that the three dimensional theory
describing the M2-brane system
should be superconformal in the infra-red, with global symmetry
enhanced from O(7) to O(8) \cite{SS,BS}.
We will discuss various of these strongly coupled configurations. 

Finally, in section 6 we consider non-abelian D$p$ brane-antibrane systems,
and show that the polarized branes discussed previously  
arise as worldvolume solitons from these systems after tachyon condensation.
As we will see,
the non-abelian brane-antibrane system couples to the right terms 
to describe this situation. In section 7
we identify the corresponding M-theory brane-antibrane configurations
and show in section 8 that the reduction to Type IIA predicts as well
the appearance
of non-abelian expanding F1-branes, 
NS5-branes and Kaluza-Klein monopoles as topological solitons.

\section{The non-abelian action}

In this section we summarize the main results in \cite{Myers}
and \cite{TvR}. We will be using the notation of \cite{Myers}.
The reader is referred to these references for more details.

The Born-Infeld part of the worldvolume effective action proposed
in \cite{Myers,TvR} to describe a system of N coincident 
D$p$-branes is given by:

\begin{equation}
\label{nonabDBI}
S_{\rm BI}=-T_p\int {\rm Tr}\left( e^{-\phi}
\sqrt{-{\rm det}(P[E_{ab}+E_{ai}(Q^{-1}-\delta)^{ij}E_{jb}]
+(2\pi\alpha^\prime)F_{ab}){\rm det}(Q^i_j)}\right)\, .
\end{equation}

\noindent Here
$Q^i_j\equiv\delta^i_j+i(2\pi\alpha^\prime)[\Phi^i,\Phi^k]E_{kj}$,
$E_{\mu\nu}\equiv g_{\mu\nu}+B_{\mu\nu}^{(2)}$, 
and $g_{\mu\nu}$, $B_{\mu\nu}^{(2)}$ are
the ten dimensional spacetime metric and NS-NS 2-form. 
Static gauge is assumed, i.e. the worldvolume coordinates are taken as 
$\xi^a=X^a$ for $a=0,1,\dots ,p$, whereas  
the remaining spacetime coordinates are rescaled as 
$X^i=2\pi\alpha^\prime \Phi^i$, with $i=p+1,\dots ,9$, in such a way
that the adjoint scalars
$\Phi^i$ have dimensions of ${\rm length}^{-1}$, like the gauge fields. 
$F_{ab}$ is
the non-abelian Born-Infeld field strength:
$F_{ab}=\partial_a A_b-\partial_b A_a +i[A_a, A_b]$, with 
$A_a=A_a^{(n)}T_n$ and $T_n$  
the ${\rm N}^2$ generators of U(N), normalized such that 
${\rm Tr}(T_nT_m)={\rm N}\delta_{nm}$. Finally, $P$ denotes the 
pull-back to the $p+1$ dimensional worldvolume, 
which in the non-abelian case is defined with
covariant derivatives \cite{Hull}
${\cal D}_a \Phi^i=\partial_a \Phi^i+i[A_a, \Phi^i]$, as 
required by gauge invariance and implied by T-duality. 
In this action the
symmetrized trace prescription of Tseytlin \cite{Tseytlin} is adopted
(see however \cite{eric} and references therein).

The WZ part of the action reads \cite{Myers,TvR}:

\begin{equation}
\label{nonabWZ}
S_{\rm WZ}=\mu_p\int {\rm Tr}\left( P\left[e^{i(2\pi\alpha^\prime)
i_\Phi i_\Phi}(\sum_n C^{(n)}e^{B^{(2)}})\right]
e^{2\pi\alpha^\prime F}\right)\, .
\end{equation}

\noindent Here $i_\Phi$ denotes the interior product with $\Phi^i$:
$(i_\Phi C^{(r)})_{i_2\dots i_r}=\Phi^{i_1}
C^{(r)}_{i_1\dots i_r}$. The T-duality
analysis reveals that it must act both on the NS-NS 2-form 
and the RR potentials. 
The most striking aspect of this action is that it involves couplings to 
RR potentials
with form degree larger than the dimensionality of the worldvolume.
These fields can couple to the $p+1$ dimensional worldvolume by
means of the
interior products with the non-abelian scalars. The presence of these
couplings has been confirmed by direct examination of string
scattering amplitudes in \cite{GM2}.
Some implications were already analyzed in \cite{Myers} 
and \cite{CMT}, where it was 
shown that they gave charge to interesting 
non-commutative solutions in the worldvolume of the N D$p$-branes. 

Let us consider as an illustration the collection 
of N D0-branes in a constant RR 4-form field strength discussed
in \cite{Myers}.
The coupling of the N D0-branes to this field can be read from 
(\ref{nonabWZ}), in particular from:

\begin{equation}
\label{nonabD0}
\int {\rm Tr}P\left[i_\Phi i_\Phi C^{(3)}\right]=
\int dt {\rm Tr}\left[\Phi^j \Phi^i (C^{(3)}_{ijt}(\Phi,t)+
(2\pi\alpha^\prime)C^{(3)}_{ijk}(\Phi,t){\cal D}_t\Phi^k)\right]\, .
\end{equation}

\noindent Since the background fields are functionals of the 
non-abelian scalars
$C^{(3)}(\Phi,t)$ is defined in terms of the non-abelian 
Taylor expansion \cite{GM}:

\begin{equation}
\label{expan}
C^{(3)}(\Phi,t)=e^{(2\pi\alpha^\prime)\Phi^i\partial_{x^i}}C^{(3)}(t)=
C^{(3)}(t)+(2\pi\alpha^\prime)\Phi^k\partial_k C^{(3)}(t)+\dots\, .
\end{equation}

\noindent Contraction of the first term in (\ref{expan})
with $i_\Phi i_\Phi$ gives a 
vanishing trace, whereas the second term together with the first 
contribution to
the expansion of the last term in
(\ref{nonabD0}) yield:

\begin{equation}
\int dt {\rm Tr} (\Phi^i \Phi^j \Phi^k)F^{(4)}_{tijk}(t)\, .
\end{equation}

\noindent Combining this term with the leading order
scalar potential from the Born-Infeld action it is possible to construct
an explicit
non-commutative solution to the equations of motion with a non-vanishing
dipole coupling \cite{Myers}. One can also notice that going beyond 
leading order a whole series
of higher order multipole couplings arises.
This is the D-brane analog of the dielectric effect of
electromagnetism. The nontrivial $F^{(4)}$ field has the effect of polarizing
the D0-branes to expand into a non-commutative configuration which can be
interpreted as a spherical D2-brane with N D0-branes bound to it.
Reference \cite{Myers} also investigated to what extent 
this configuration could be described as a solution
in the abelian worldvolume theory of the D2-brane, with the remarkable result 
that the two approaches
agree up to $1/{\rm N}^2$ corrections.

The previous analysis can be generalized to D$p$-branes in a
background of constant $F^{(p+4)}$ field strength. 
Starting with a flat D$p$-brane
it is energetically favorable for the brane to expand into a 
non-commutative configuration with spatial geometry $R^p \times S^2$ and
non-vanishing dipole charge, as implied by the term:

\begin{equation}
\int d{\xi}^0 d{\xi}^1\dots d{\xi}^p {\rm Tr}(\Phi^i \Phi^j \Phi^k)
F^{(p+4)}_{01\dots p ijk}(\xi)\, .
\end{equation}

\noindent The contribution of this term to the scalar
potential for N coincident D-strings was discussed in \cite{CMT},
and it was shown that there exists a
non-commutative solution describing polarized D-strings, which is interpreted
as a spherical (N D1, D3) bound state. 
Reference \cite{TV} considered a more general $(p+4)$-form field
strength than the one considered in \cite{Myers} and obtained other
non-commutative configurations corresponding to certain fuzzy cosets.

As another interesting application reference
\cite{CMT} considered a similar type of non commutative solution 
in the worldvolume of N coincident D-strings in the absence of any nontrivial
background fields. This solution was interpreted as a funnel
describing the expansion of the D-strings into an orthogonal D3-brane.
This configuration  
acts as a source for the RR 4-form potential thanks to the
non-trivial expectation values of the non-abelian scalars, which
contribute through the term:
$\int_{R^{1+1}}{\rm Tr}(i_\Phi i_\Phi C^{(4)})$.
The dual description in the worldvolume of the 
D3-brane is in terms of a spike
solution corresponding to D-strings attached to
the D3-brane, configuration that has been extensively studied in 
the literature \cite{CM}. In
\cite{CMT} it is shown that the two descriptions have complementary 
ranges of validity, 
so that the non-commutative D-string theory point of view
is reliable at the center of the
spike, where the D3-brane description is expected to break down. Again both
descriptions turn out to be in agreement in the large N limit. 
As pointed out in \cite{CMT} the extension to D$p$-branes opening up
into orthogonal D$(p+2)$-branes is straightforward. It is
governed by the term:
$\int_{R^{p+1}}{\rm Tr}(i_\Phi i_\Phi C^{(p+3)})$.

In the next section we will discuss similar type of non-commutative
couplings in M-theory.

\section{Non-abelian configurations in M-theory}

Let us start by analyzing the kind of non-abelian configurations of
branes that one can construct in M-theory. Our approach will be to
represent these non-abelian systems as Kaluza-Klein monopoles
(or M9-branes) with N units of worldvolume instanton-like density.

The worldvolume field content of the M-theory Kaluza-Klein
monopole is that
of the seven dimensional U(1) vector multiplet, involving 3 scalars
and 1 vector\footnote{Recall that the embedding coordinates contribute
with 3 degrees of freedom because one of the scalars is eliminated 
through the gauging of the Taub-NUT isometry of the background 
\cite{BJO}.}.
A system of N coincident
Kaluza-Klein monopoles is then described by a seven dimensional
U(N) vector multiplet.
Therefore the Wess-Zumino action contains the couplings 
(see \cite{BEL})\footnote{We
ignore all numerical prefactors and the contribution of the A-roof genus.
Hats indicate eleven dimensional fields. $l_p$ denotes the eleven
dimensional Planck length. Our conventions are that:
$(i_{\hat k}{\hat L})_{{\hat \mu}_2\dots{\hat \mu}_r}=
{\hat k}^{{\hat \mu}_1} {\hat L}_{{\hat \mu}_1\dots{\hat \mu}_r}$,
and $i_{\hat k}{\hat N}^{(8)}=C^{(7)}$, $i_{\hat k}{\hat {\tilde C}}=
C^{(5)}+\dots$ upon reduction along the Killing direction.}:

\begin{equation}
\label{Mmono}
S^{N KK}_{WZ}=\mu_6\int_{R^{6+1}}{\rm Tr}\left( i_{\hat k}{\hat N}^{(8)}
+l_p^2 i_{\hat k}{\hat {\tilde C}}\wedge {\hat {\cal F}}+l_p^4
{\hat C}_{{\hat \mu}{\hat \nu}{\hat \rho}}D{\hat X}^{\hat \mu}
D{\hat X}^{\hat \nu}D{\hat X}^{\hat \rho}\wedge {\hat {\cal F}}\wedge
{\hat {\cal F}}+\dots\right)\, .
\end{equation}

\noindent ${\hat k}^{\hat \mu}$ denotes the Killing vector along the
Taub-NUT direction, and
${\hat N}^{(8)}$ is its Poincar\'e dual when
considered as a 1-form ${\hat k}_{\hat \mu}$.
${\hat C}$ (${\hat {\tilde C}}$) denote the 3-form
(6-form) of eleven dimensional supergravity.  
The D-derivatives are defined as:
$D{\hat X}^{{\hat \mu}}={\cal D}{\hat X}^{{\hat \mu}}-
{\hat k}^{-2}{\hat k}_{{\hat \nu}}{\cal D}{\hat X}^{{\hat \nu}}
{\hat k}^{{\hat \mu}}$, with
${\hat k}^2={\hat g}_{{\hat \mu}{\hat \nu}}
{\hat k}^{\hat \mu}{\hat k}^{\hat \nu}$. Note that covariant
derivatives substitute partial derivatives in the non-abelian case,
in such a way that:  
${\hat C}D{\hat X}D{\hat X}D{\hat X}=
\left({\hat C}-3{\hat k}^{-2}
{\hat k}^{(1)} i_{\hat k}{\hat C}\right)
{\cal D}{\hat X}{\cal D}{\hat X}{\cal D}{\hat X}$, 
and the contribution of the 
Taub-NUT direction cancels out\footnote{Here ${\hat k}^{(1)}$ denotes
the Killing vector considered as a 1-form, with components
${\hat k}_{\hat \mu}$.}. 
${\hat {\cal F}}$ is the
field strength of the U(N) vector field describing 
M2-branes, wrapped in the Killing direction\footnote{This is implied
by the fact that ${\hat C}$ is contracted with the Killing vector
\cite{BEL}.}, ending on the monopoles:
${\hat {\cal F}}=2\partial{\hat A}+i[{\hat A},{\hat A}]+
l_p^{-2}i_{\hat k}{\hat C}
\equiv {\hat F}+l_p^{-2}i_{\hat k}{\hat C}$. The spacetime fields
are understood to be pulled-back onto the worldvolume as explained
in section 2.

The second term in (\ref{Mmono}) shows that wrapped M5-branes can
arise as solitonic solutions when there is a non-trivial magnetic
flux in $R^2$. Similarly, the third term shows that an instanton-like
configuration $\int_{R^4}{\rm Tr}{\hat F}\wedge {\hat F}=Z$ induces
M2-brane charge, but with the M2-branes delocalized along
the Killing direction, since the resulting
coupling contains $D{\hat X}$ derivatives:
$\int_{R^{2+1}}{\hat C}D{\hat X}D{\hat X}D{\hat X}$.
 
The same situation in terms of wrapped M5-branes and delocalized
M2-branes arises from the analysis of the Wess-Zumino action of
a system of M9-branes. The field content of the M9-brane is that
of the nine dimensional U(1) vector multiplet, containing 1 scalar
and 1 vector. The vector contributes however with 7 degrees of
freedom when one of the worldvolume directions is gauged 
away\footnote{The Killing vector points at a worldvolume direction,
in such a way that the D8-brane is obtained upon reduction.}
(see \cite{BvdS}). Therefore, a system of coincident N M9-branes
is described by a nine dimensional U(N) vector multiplet.
The Wess-Zumino action contains the terms \cite{Sato}:

\begin{eqnarray}
\label{NM9}
S^{N M9}_{WZ}&=&\mu_8\int_{R^{8+1}}{\rm Tr}\left(i_{\hat k}{\hat B}^{(10)}
+l_p^2 i_{\hat k}{\hat N}^{(8)}\wedge {\hat {\cal F}}
+l_p^4 i_{\hat k}{\hat {\tilde C}}
\wedge {\hat {\cal F}}\wedge {\hat {\cal F}}+\right. \nonumber\\
&&\left. +l_p^6 {\hat C}D{\hat X}D{\hat X}D{\hat X}\wedge 
{\hat {\cal F}}
\wedge {\hat {\cal F}}\wedge {\hat {\cal F}}+\dots\right)\, .
\end{eqnarray}

\noindent {}From here we see that
Kaluza-Klein monopole charge is induced when 
$\int_{R^2}{\rm Tr}{\hat F}=Z$, (wrapped) M5-brane charge when
$\int_{R^4}{\rm Tr}({\hat F}\wedge {\hat F})=Z$, and (delocalized)
M2-brane charge when $\int_{R^6}{\rm Tr}({\hat F}\wedge{\hat F}
\wedge {\hat F})=Z$.

The analysis of the brane descent relations of M-theory points
towards the same situation. It is possible to construct
brane descent relations in M-theory from non-abelian
configurations of non-BPS M10-branes. The M10-brane is constructed in
such a way that it gives rise to the non-BPS
D9-brane of Type IIA \cite{Horava} after 
reduction along a Killing direction, and to the BPS M9-brane after
condensation of its tachyonic mode. This connection with the M9-brane
determines that it should contain a Killing direction in its 
worldvolume. The analysis of the WZ couplings reveals the
following pattern of brane descent relations: M9=M10, M6=2 M10, 
M5=4 M10, M2=8 M10, M0=16 M10, where M6 denotes the Kaluza-Klein
monopole and M0 the M-wave. This analysis was made in \cite{HL3}.
There it was pointed out that the M5-brane should again be wrapped
around the Killing direction and the M2-brane delocalized in this
direction. This is implied by the following terms
in the M10-brane effective action:

\begin{equation}
\label{M10}
S^{N M10}_{{\rm WZ}}=\mu_9\int_{R^{9+1}}{\rm Tr}\left(\left[l_p^4
i_{\hat k}
{\hat {\tilde C}}
\wedge {\hat {\cal F}}\wedge {\hat {\cal F}}+
l_p^6{\hat C}_{{\hat \mu}
{\hat \nu}{\hat \rho}}D{\hat X}^{\hat \mu}
D{\hat X}^{\hat \nu}D{\hat X}^{\hat \rho}\wedge 
{\hat {\cal F}}
\wedge {\hat {\cal F}}\wedge {\hat {\cal F}}
+\dots\right]\wedge {\cal D}{\hat T}\right)\, ,
\end{equation}

\noindent where ${\hat T}$ stands for
the real adjoint tachyon induced in the worldvolume by wrapped
M2-branes,
and ${\cal D}{\hat T}=d{\hat T}+i[{\hat A},{\hat T}]$\footnote{The 
full Wess-Zumino action for coincident 
non-BPS D-branes has been constructed recently in \cite{JM}, 
extending previous results in \cite{Kluson,BCR}. In this reference
it is shown
that this action contains an infinite sum
of terms with different powers of the tachyon and its covariant 
derivatives. Similar kind of terms will also couple in the
worldvolume effective action describing non-abelian M10-branes.
For our purposes it will be sufficient to just consider the
contribution of the previous term in this expansion, from where the
desired couplings can already be read. In (\ref{M10})
the condensation of the tachyon through a kink-like configuration,
which in the limit of zero size can be written as:
$d{\hat T}={\hat T}_0 \delta (x-x_0) dx$ \cite{BCR},
gives rise to the Wess-Zumino term of a system of N M9-branes
located at $x=x_0$.}.

The analysis of the brane descent relations points out 
that arbitrary M2 and M5 branes cannot be connected to the other
branes of M-theory through a hierarchy of embeddings. The fact that
the higher dimensional M-branes live in spacetimes
with special Killing directions implies that
the branes that can be constructed from them as bound states
should also see the special direction. 
The analysis of the Wess-Zumino
terms of the higher dimensional branes shows in particular that the
M5-brane should be wrapped 
and the M2-brane should not move along the Killing direction.
In this situation
one can construct non-abelian configurations
of M5 and M2 branes as bound states of 4N M10-branes and 8N M10-branes
respectively.
We can conclude that the M5-brane 
that arises as a bound state in a system of
higher dimensional branes behaves
effectively as a 4-brane propagating in a ten dimensional spacetime. 
Thus, its field content must
be that of the five dimensional vector multiplet, whose non-abelian
extension is well-known. Similarly, the M2-branes behave like 
2-branes in ten dimensions, and therefore their field content should
be that of the three dimensional U(M) vector multiplet.

Moreover, the existence of the expanded configurations discussed
in the previous section in
the Type IIA theory implies that it should be possible to construct
configurations in M-theory
corresponding to M2-branes or M5-branes
expanding into higher dimensional branes. However, the fact that
the higher dimensional branes are coupled to spacetime
fields that are constrained by the presence of the Killing direction  
implies that the non-abelian worldvolume theory that would describe 
the M2 and M5 brane
configurations should also be constrained by the existence of the 
Killing direction, which is consistent with our discussion above.
This fact does not exclude however the possibility of 
constructing non-abelian configurations of arbitrary
M2-branes opening up into unwrapped M5-branes\footnote{These 
branes will however not be able to expand into other higher 
dimensional branes such as the monopole and the
M9-brane, since they cannot see their special isometric directions.}.
The existence
of this kind of configuration is in fact predicted by the coupling
$\int_{R^{5+1}} {\hat C}\wedge d{\hat a}^{(2)}$ present in the
worldvolume effective action of a single M5-brane. The difficulty stands
however in the construction of the non-abelian worldvolume theory 
that describes
the set of coinciding M2-branes. This problem is related to the problem of
implementing duality transformations in non-abelian gauge
theories, as we will further discuss in the paper.

Let us now discuss which could then be the
M-theory description of a system of
N coincident D2-branes. The Wess-Zumino action of this Type IIA system,
up to linear terms in the non-abelian Born-Infeld field strength, 
includes \cite{Myers}:

\begin{eqnarray}
\label{actionND2}
&&S^{ND2}_{WZ}=\mu_2\int_{R^{2+1}}{\rm Tr}\left(C^{(3)}+
i(2\pi\alpha^\prime)i_\Phi i_\Phi C^{(5)}-\frac12 (2\pi\alpha^\prime)^2
(i_\Phi i_\Phi)^2 C^{(7)}+\dots\right.\nonumber\\
&&\left.
+(2\pi\alpha^\prime)(C^{(1)}+i(2\pi\alpha^\prime)i_\Phi i_\Phi C^{(3)}
-\frac12 (2\pi\alpha^\prime)^2 (i_\Phi i_\Phi)^2 C^{(5)}+\dots)
\wedge {\cal F}+\dots\right)\, ,
\end{eqnarray}

\noindent where ${\cal F}=F+\frac{1}{2\pi\alpha^\prime}B^{(2)}$.
From our discussion in the previous section the interpretation
of the couplings in the first line of (\ref{actionND2}) should be clear. 
The contraction of the embedding scalars with the RR 5-form 
indicates the existence of a non-commutative configuration 
corresponding to the D2-branes expanding into a D4-brane, which acts
as a source for this RR potential, and the next term:
$(i_\Phi i_\Phi)^2 C^{(7)}$, should represent the N D2-branes
expanding into a D6-brane\footnote{See the discussions in \cite{CMT} for 
the similar configuration of D-strings opening up into a D5-brane,
and \cite{GJS} for a solution associated to D-instantons in a RR
5-form field strength. \cite{TV} also considers different four
dimensional non-commutative configurations.}.
 
Uplifting these couplings to M-theory one finds\footnote{We have
denoted these branes as $M2_t$ to specify that they are transverse to the
Killing direction.}:

\begin{eqnarray}
\label{NM2}
&&S^{NM2_t}_{WZ}=\mu_2\int_{R^{2+1}}{\rm Tr}\left({\hat C}D{\hat X}D{\hat X}
D{\hat X}+il_p^2 i_{\hat \Phi} i_{\hat \Phi} i_{\hat k} 
{\hat {\tilde C}}-\frac12 l_p^4 (i_{\hat \Phi} i_{\hat \Phi})^2
i_{\hat k}{\hat N}^{(8)}+\dots\right.\nonumber\\
&&\left.+l_p^2({\hat k}^{-2}{\hat k}^{(1)}+il_p^2 i_{\hat \Phi}i_{\hat \Phi}
({\hat C}-{\hat k}^{-2}{\hat k}^{(1)}\wedge i_{\hat k}{\hat C})
-\frac12 l_p^4 (i_{\hat \Phi}i_{\hat \Phi})^2 i_{\hat k}{\hat {\tilde C}}
+\dots)\wedge {\hat {\cal F}}+\dots\right)\, ,
\end{eqnarray}

\noindent with ${\hat {\cal F}}=2\partial {\hat A}+
i[{\hat A},{\hat A}]+l_p^{-2}i_{\hat k}{\hat C}$.
This action describes a non-abelian configuration 
of M2-branes delocalized along the eleventh direction,
which appears as a special Killing direction mainly for two reasons.
First, in the non-abelian case it is not possible to dualize the
Born-Infeld field of the D2-branes into the scalar associated with
the eleventh coordinate,
a necessary step in order to connect the 
fully eleven dimensional M2-brane and the D2-brane. Therefore, we
are constrained to introduce the eleventh direction as a special
isometric direction. 
Second, the spacetime fields
contracted with the embedding scalars cannot be uplifted into
any, unwrapped, eleven dimensional field.
Thus, (\ref{NM2}) is describing the same type of
M2-branes that can arise as worldvolume solitons in a single
M6 or M9 brane. 
The second coupling in (\ref{NM2}) describes a non-commutative
configuration corresponding to the N M2-branes expanding into a wrapped
M5-brane. This configuration has been considered in \cite{Bena2} in
the context of the AdS/CFT correspondence. Here we find that it is
possible to describe it from the point of view of the non-abelian
system of branes if the M2-branes are delocalized and the M5-branes
are wrapped around the eleventh direction.
With respect to the interpretation of the third coupling,
we have seen that $i_{\hat k} {\hat N}^{(8)}$ is the field to which
the M-theory Kaluza-Klein monopole couples minimally\footnote{It
appears in the N M2-branes effective action because it 
reduces to the term $(i_\Phi i_\Phi)^2 C^{(7)}$ in (\ref{actionND2})
(see \cite{BEL}).}. Therefore its contraction with the four 
non-commutative scalars would give charge to
a configuration corresponding to the N M2-branes expanding into
a Kaluza-Klein monopole. The existence of this configuration also
explains why the M2-branes contain a special
Killing direction. We will give an interpretation of the terms
in the second line of (\ref{NM2}) in section 7, where we analyze
solitonic configurations in non-abelian brane-antibrane systems.

Similarly, the Wess-Zumino part of the worldvolume effective action
describing a system of N M5-branes is constructed from that of
N D4-branes. This includes the terms:

\begin{eqnarray}
\label{actionND4}
&&S_{{\rm WZ}}^{{\rm N D4}}=\mu_4 \int_{R^{4+1}}{\rm Tr}
\left(C^{(5)}+i(2\pi\alpha^\prime)
i_\Phi i_\Phi C^{(7)}-\frac12 (2\pi\alpha^\prime)^2 (i_\Phi i_\Phi)^2
C^{(9)}+\right.\nonumber\\
&&\left. +(2\pi\alpha^\prime)(C^{(3)}+
i(2\pi\alpha^\prime)i_\Phi i_\Phi
C^{(5)}-\frac12 (2\pi\alpha^\prime)^2 (i_\Phi i_\Phi)^2 C^{(7)}+\dots)
\wedge {\cal F}+\dots\right)\, .
\end{eqnarray}

\noindent The contraction of the embedding scalars with the RR 7-form 
indicates the existence of a configuration corresponding to the D4-branes 
expanding into a D6-brane, whereas the
term $(i_\Phi i_\Phi)^2 C^{(9)}$, would be associated to the N D4-branes
expanding into a D8-brane. Uplifting this non-abelian system onto
M-theory we find a non-abelian configuration of wrapped M5-branes
containing the couplings:

\begin{eqnarray}
\label{actionNM5}
&&S^{N {M5}_w}_{\rm WZ}=\mu_4\int_{R^{4+1}}{\rm Tr}\left(i_{\hat k} 
{\hat {\tilde C}}+il_p^2 i_{\hat \Phi} i_{\hat \Phi} i_{\hat k} 
{\hat N}^{(8)}-\frac12 l_p^4
(i_{\hat \Phi} i_{\hat \Phi})^2 i_{\hat k}{\hat B}^{(10)}+
\right.\nonumber\\
&&+\left. l_p^2({\hat C}D{\hat X}D{\hat X}D{\hat X}+il_p^2
i_{\hat \Phi} i_{\hat \Phi} i_{\hat k}{\hat {\tilde C}}-\frac12 l_p^4
(i_{\hat \Phi} i_{\hat \Phi})^2 i_{\hat k}{\hat N}^{(8)}+\dots)\wedge
{\hat {\cal F}}+\dots\right)\, .
\end{eqnarray}

\noindent The Killing direction emerges, on the one hand,
because the vector field of the D4-branes cannot be (worldvolume)
dualized into the worldvolume 2-form of an unwrapped
M5-brane and, on the other
hand, because it is not possible to uplift the fields contracted with
the embedding scalars into any eleven dimensional fields not 
involving the Killing vector.
The contraction of $i_{\hat k}{\hat N}^{(8)}$ with the
two non-commutative scalars gives charge to a configuration
corresponding to the N M5-branes opening up into a Kaluza-Klein
monopole. The third coupling gives charge in turn to a configuration
corresponding to the N M5-branes opening up into an M9-brane, since
as we have seen the field $i_{\hat k}{\hat B}^{(10)}$ couples
minimally to this brane\footnote{It 
gives the 9-form RR-potential
of Type IIA after reduction along the Killing direction.}.
Similarly to the
M2-brane case, the M5-branes must be wrapped on the Killing 
direction, consistently with the fact that they should
be able to expand into the monopole or the M9-brane.

The Wess-Zumino action describing a non-abelian configuration 
of Kaluza-Klein monopoles is obtained from that of a system of N 
D6-branes, following the same procedure described in \cite{BEL}.
One obtains the action (\ref{Mmono}) supplemented with typically
non-commutative
couplings corresponding to the contraction of the spacetime fields
with the embedding scalars:

\begin{eqnarray}
\label{Mmono2}
S^{NKK}_{WZ}&=&\mu_6 \int_{R^{6+1}}{\rm Tr}\left(i_{\hat k}{\hat N}^{(8)}
+il_p^2 i_{\hat \Phi}i_{\hat \Phi}i_{\hat k}{\hat B}^{(10)}
+l_p^2(i_{\hat k}{\hat {\tilde C}}+il_p^2 i_{\hat \Phi}
i_{\hat \Phi}i_{\hat k}{\hat N}^{(8)}-\right.\nonumber\\
&&\left. -\frac12 l_p^4 (i_{\hat \Phi}
i_{\hat \Phi})^2 i_{\hat k} {\hat B}^{(10)})\wedge {\hat {\cal F}}
+\dots\right)\, .
\end{eqnarray}

\noindent The second term in (\ref{Mmono2}) can be interpreted in
terms of a configuration of N monopoles opening up into an
M9-brane.

Finally, the non-abelian action associated to coincident M-waves
is obtained by uplifting the Wess-Zumino action of a system of
D0-branes. One finds the couplings:

\begin{equation}
S^{N M0}_{WZ}=\mu_0\int_R {\rm Tr}\left({\hat k}^{-2}{\hat k}^{(1)}
+il_p^2 i_{\hat \Phi}i_{\hat \Phi}({\hat C}-{\hat k}^{-2}{\hat k}^{(1)}
\wedge i_{\hat k}{\hat C})-\frac12 l_p^4 
(i_{\hat \Phi} i_{\hat \Phi})^2 i_{\hat k}{\hat {\tilde C}}
+\dots\right)\, .
\end{equation}

\noindent The second term indicates the existence of a configuration
associated to M-waves expanding into an M2-brane transverse to the 
direction of propagation. Therefore,
this is the coupling that is responsible for the `dual
giant graviton' of $AdS_4\times S^7$ \cite{GMT}.
In turn the third term represents the N M0-branes expanding into an
M5-brane wrapped in the direction of the propagation, and is associated
to the dual giant graviton of $AdS_7\times S^4$. 
These configurations can also be studied
from the point of view of the M2 and M5 branes. 
The first term in the
second line of (\ref{NM2}) indicates that the transverse M2-brane
carries momentum in a direction orthogonal to the electric 3-form.
In the case of the wrapped M5-brane the term:

\begin{equation}
\int_{R^{4+1}} {\rm Tr} [{\hat k}^{-2}{\hat k}^{(1)}\wedge {\hat {\cal F}}
\wedge {\hat {\cal F}}]\, ,
\end{equation}

\noindent that we have omitted in (\ref{actionNM5}), shows that it
carries momentum along the compact direction and that this arises as 
a non-trivial instanton charge. 

Let us finish this section with some comments on the M2 and M5 brane
effective actions that we have presented.
In the abelian case the action of the $M2_t$-brane would be related to
that of an ordinary M2-brane by means of a worldvolume duality
transformation. This is easily seen by 
considering the abelian version of the action (\ref{NM2})
and adding to it a Lagrange multiplier term
$\int_{R^{2+1}} d{\hat A}\wedge d{\hat y}=\int_{R^{2+1}}
({\hat {\cal F}}-l_p^{-2}
i_{\hat k}{\hat C})\wedge d{\hat y}$. The integration
on ${\hat y}$ would impose
the constraint that ${\hat {\cal F}}-l_p^{-2}
i_{\hat k}{\hat C}$ derived from a vector potential
and the original action would be recovered. On the other hand, the
integration of ${\hat {\cal F}}$ would give rise to a coupling:
$\int_{R^{2+1}} [{\hat C}+i_{\hat k}{\hat C}\wedge d{\hat y}]$
in the dual action. This term describes an M2-brane, with
${\hat y}$ playing the role of the eleventh direction. 
This connection between $M2_t$ and M2 does not exist however in the
non-abelian case and only systems involving the first type of branes
can be constructed explicitly.
The $M5_w$-brane that we have constructed
can also be related to an ordinary M5-brane in the abelian case.
The worldvolume duality transformation would be performed
by adding the Lagrange multiplier term:
$\int_{R^{4+1}} d{\hat A}\wedge d{\hat a}^{(2)}=\int_{R^{4+1}}
({\hat {\cal F}}-l_p^{-2}i_{\hat k}{\hat C})\wedge d{\hat a}^{(2)}$
to the action (\ref{actionNM5}).
Now a coupling $\int_{R^{4+1}}i_{\hat k}
{\hat C}\wedge d{\hat a}^{(2)}$ would remain after integrating out
${\hat {\cal F}}$, which would arise in the double
dimensional reduction of an M5-brane.

\section{Other non-abelian Type IIA configurations}

We have seen that a system of N M2-branes
can open up into a wrapped M5-brane when they are delocalized along
the direction on which the M5-brane is wrapped.
This configuration gave rise upon reduction along the special
direction to N D2-branes opening up into a D4-brane. One could
however consider other possibilities. Consider for instance reducing 
this system along a transverse direction different
from the Killing direction.
This would give rise to a configuration of N D2-branes expanding into
an NS5-brane, with the NS5-brane wrapped and the
D2-branes constrained to move in the transverse space. 
This is in fact described by the following couplings in the D2-branes
effective action:

\begin{eqnarray}
\label{actiontD2}
&&S^{ND2_t}_{WZ}=\mu_2\int_{R^{2+1}}{\rm Tr}\left(C^{(3)}DXDXDX+
i(2\pi\alpha^\prime)i_\Phi i_\Phi i_k B^{(6)}-\frac12 
(2\pi\alpha^\prime)^2 (i_\Phi i_\Phi)^2 i_k N^{(8)}+\dots
\right.\nonumber\\
&&+(2\pi\alpha^\prime)(k^{-2}k^{(1)}+i(2\pi\alpha^\prime)
i_\Phi i_\Phi (C^{(3)}DXDXDX)-\frac12 (2\pi\alpha^\prime)^2
(i_\Phi i_\Phi)^2 i_k B^{(6)}+\dots)\wedge {\cal H}^{(2)}+
\nonumber\\
&&\left.+\dots\right)\, ,
\end{eqnarray}

\noindent derived from (\ref{NM2}) upon reduction\footnote{We denote
these branes as $D2_t$ to specify that they are transverse to the
Killing direction.}. ${\cal H}^{(2)}$ is defined as:
${\cal H}^{(2)}=2\partial b^{(1)}+i[b^{(1)},b^{(1)}]+
\frac{1}{2\pi\alpha^\prime}i_k C^{(3)}
+2 (i_k B^{(2)}){\cal D}c^{(0)}$. Here $c^{(0)}$ is
the scalar field arising from the eleventh direction:
$c^{(0)}=y/(2\pi\alpha^\prime)$ and $b^{(1)}$ comes from the
reduction of the vector field, which we have denoted as $b^{(1)}$
to specify that its abelian part has different gauge transformation
rule than the vector field $A$ that couples in ordinary D$p$-brane
effective actions. We have also taken $C^{(1)}=0$ for 
simplicity\footnote{We will be doing the same in the coming actions.}.
${\cal H}^{(2)}$ is associated to wrapped D2-branes ending on the
non-abelian system of branes.
Before we discuss the interpretation
of the different couplings in this effective action let us consider
other possibilities.

We could also reduce the N M2-branes along a
wordvolume direction. This would give rise to N fundamental strings,
constrained to move in a nine dimensional spacetime, which could
expand into a
wrapped D4-brane. The corresponding
couplings that one finds after the reduction are:

\begin{eqnarray}
\label{actiontF1}
&&S^{NF1_t}_{WZ}=\mu_1\int_{R^{1+1}}{\rm Tr}\left(B^{(2)}DXDX+
i(2\pi\alpha^\prime)i_\Phi i_\Phi i_k C^{(5)}-\frac12 
(2\pi\alpha^\prime)^2 (i_\Phi i_\Phi)^2 i_k N^{(7)}+\dots
\right.\nonumber\\
&&+(2\pi\alpha^\prime)(k^{-2}k^{(1)}+i(2\pi\alpha^\prime)
i_\Phi i_\Phi (C^{(3)}DXDXDX)-\frac12 (2\pi\alpha^\prime)^2
(i_\Phi i_\Phi)^2 i_k B^{(6)}+\dots)\wedge {\cal K}^{(1)}+\nonumber\\
&&\left. +(2\pi\alpha^\prime)(i(2\pi\alpha^\prime)
i_\Phi i_\Phi (B^{(2)}DXDX)-\frac12 (2\pi\alpha^\prime)^2
(i_\Phi i_\Phi)^2 i_k C^{(5)}+\dots)\wedge {\cal K}^{(2)}
+\dots\right)\, ,\nonumber\\
\end{eqnarray}

\noindent where ${\cal K}^{(1)}={\cal D}\omega^{(0)}+
\frac{1}{2\pi\alpha^\prime}(i_k B^{(2)})+i(2\pi\alpha^\prime)
(i_k C^{(3)}){\cal D}\Phi [\omega^{(0)},\Phi]$ and
${\cal K}^{(2)}=2\partial\omega^{(1)}+i[\omega^{(1)},\omega^{(1)}]
+\frac{1}{2\pi\alpha^\prime}i_k C^{(3)}$. $\omega^{(0)}$ and
$\omega^{(1)}$ arise as the components of ${\hat A}$ along the direction
in which we reduce and along a different direction, and are
associated to wrapped F1-branes and wrapped D2-branes ending on the
F1-branes. Note that the coupling to the wrapped D2-branes only
occurs in the non-abelian case, since the spacetime fields must be
contracted with the embedding scalars. Non-abelian configurations of
Type IIA fundamental strings have been studied in \cite{ES}.

Similarly, the case in which N wrapped M5-branes opened up 
into a Kaluza-Klein monopole gave rise to N D4-branes opening up into 
a D6-brane when reducing along the Killing direction. Reduction
along a transverse
direction would give rise instead to N wrapped NS5-branes, 
containing the couplings:

\begin{eqnarray}
\label{actionNS5}
&&S^{N (NS5)_w}_{WZ}=\mu_4\int_{R^{4+1}}{\rm Tr}\left(i_k B^{(6)}+
i(2\pi\alpha^\prime)i_\Phi i_\Phi i_k N^{(8)}-\frac12
(2\pi\alpha^\prime)^2 (i_\Phi i_\Phi)^2 i_k B^{(10)}+\right.
\nonumber\\
&&\left. +(2\pi\alpha^\prime)(C^{(3)}DXDXDX+i(2\pi\alpha^\prime)
i_\Phi i_\Phi i_k B^{(6)}-\frac12 (2\pi\alpha^\prime)^2
(i_\Phi i_\Phi)^2 i_k N^{(8)}+\dots)\wedge {\cal H}^{(2)}+\dots\right)\, .
\nonumber\\
\end{eqnarray}

\noindent The second term indicates the existence of a configuration 
corresponding to the N wrapped NS5-branes expanding into a 
KK6-brane\footnote{The KK6-brane arises when reducing the M-theory
Kaluza-Klein monopole along a transverse direction different from
the Taub-NUT direction \cite{EB,HMO}, and it couples minimally
to $i_k N^{(8)}$, obtained from the reduction of 
$i_{\hat k}{\hat N}^{(8)}$. This brane is
however not predicted by the spacetime supersymmetry algebra, and 
in this sense is referred to as an exotic brane. See however the
recent work \cite{LO} for a possible extension of the SUSY algebra
including this, and other exotic, charges.}. 
Reduction along a worldvolume direction of the M5's different from 
that on which they are wrapped would give rise to
N wrapped D4-branes, containing the couplings:

\begin{eqnarray}
\label{actionD4w}
&&S^{ND4_w}_{WZ}=\mu_3\int_{R^{3+1}}{\rm Tr}\left(i_k C^{(5)}+
i(2\pi\alpha^\prime) i_\Phi i_\Phi i_k N^{(7)}-\frac12
(2\pi\alpha^\prime)^2 (i_\Phi i_\Phi)^2 i_k N^{(9)}+\right.
\nonumber\\
&&+(2\pi\alpha^\prime)(C^{(3)}DXDXDX+i(2\pi\alpha^\prime)
i_\Phi i_\Phi i_k B^{(6)}-\frac12 (2\pi\alpha^\prime)^2 i_k N^{(8)}
+\dots)\wedge {\cal K}^{(1)}+\nonumber\\
&&\left.+(2\pi\alpha^\prime)(B^{(2)}DXDX+i(2\pi\alpha^\prime)
i_\Phi i_\Phi i_k C^{(5)}-\frac12 (2\pi\alpha^\prime)^2 
(i_\Phi i_\Phi)^2 i_k N^{(7)}+\dots)\wedge {\cal K}^{(2)}+
\dots\right)\, .\nonumber\\
\end{eqnarray}

\noindent The second term indicates the existence of a
configuration in which the N D4-branes open up into a Kaluza-Klein
monopole, given that
$i_k N^{(7)}$ is the field to which
the Type IIA monopole couples minimally\footnote{It
arises in the reduction of $i_{\hat k}{\hat N}^{(8)}$ along
a worldvolume direction (see \cite{BEL}).}. 

The first coupling in (\ref{actionD4w}) seems to imply
that the 4-branes that we have obtained after the reduction
from M-theory are just wrapped D4-branes. However, one 
notices that a system of wrapped
D4-branes would contain in its effective action a coupling 
$i_\Phi i_\Phi i_k C^{(7)}$, corresponding to the D4-branes expanding
into a wrapped D6-brane, and not the coupling $i_\Phi i_\Phi i_k N^{(7)}$
that we have found after the reduction from M-theory.
The same thing happens if we try to give a meaning to the effective 
action (\ref{actiontD2})
as associated to a delocalized D2-brane in the Type 
IIA theory. Moreover, the reduction of the field strength ${\hat {\cal F}}$
shows that the dynamics of these objects is governed by wrapped D2-branes
ending on them\footnote{Also wrapped F1-branes in the case of the 
D4-brane.}. 
Thus, the reduction from M-theory provides a
description of the D2 and D4 branes  
which is fully non-perturbative, and therefore valid in the
strong coupling regime. As we have mentioned already
the origin of this different
description at strong coupling can be traced to the fact that 
the worldvolume duality transformations that are needed in order
to prove the equivalence between the transverse or wrapped brane
and the fully ten dimensional one cannot be performed in the
non-abelian case. 
The worldvolume effective actions that we have
derived in this section must be used however to describe those
situations which are typically non-perturbative and cannot be
predicted by the weakly coupled effective actions. Indeed, the expanding
brane configurations that we have discussed, predicted by the 
contraction of the spacetime fields with the embedding scalars, cannot
be explained from the weakly coupled effective actions,
like the one associated to N D2 branes expanding into an NS5-brane,
considered in \cite{Bena}. The analysis presented in this section 
shows that this configuration can be studied from the point of view of 
the non-abelian D2-branes if they are delocalized in one spatial
direction. The NS5-brane into which they expand is then wrapped on 
this direction. 
The second coupling in (\ref{actiontD2}) shows that this system 
can carry $B^{(6)}$ charge, which is however not the case for weakly
coupled D2-branes. This special situation arises naturally from the
T-duality of a configuration of N D3-branes expanding into an NS5-brane,
that we will discuss in the next section, and is in agreement with
the Type IIA NS5-brane solution with non-trivial $C^{(3)}$ charge
that was constructed in \cite{AOS}, and considered further in
\cite{Bena}. 
The same configuration can also be represented as a single (wrapped) 
NS5 with N units of (wrapped) D2-brane magnetic flux. 
This is predicted by the first coupling
in the second line of (\ref{actionNS5}).

Other non-commutative terms show that the delocalized
D2-branes and the wrapped D4-branes can expand into higher dimensional
branes defined in isometric spacetimes, namely D2 into KK6, D4 into
a monopole, etc. which is possible because these strongly coupled
configurations can also see the special Killing direction.

We have found as well the Wess-Zumino part of the worldvolume 
effective actions describing
coincident systems of F1-branes and NS5-branes. The F1-branes
cannot move along a special Killing direction and the NS5-branes
are wrapped on it. The dynamics of these objects is governed by
wrapped D2-branes ending on them\footnote{Also wrapped F1-branes in the
case of the F1's.}. Both effective actions can be shown to reduce
to the effective action of ordinary, localized, F1-branes and
unwrapped NS5-branes in the abelian case, by means of a worldvolume
duality transformation.

Finally, there are two cases in which the reduction from M-theory
provides the only possible description of the brane. This
happens when reducing a system of N coincident Kaluza-Klein
monopoles along a worldvolume direction, which
gives rise to coinciding Type IIA monopoles. We find the couplings:

\begin{eqnarray}
\label{monoIIA}
&&S^{NKK}_{WZ}=\mu_5\int_{R^{5+1}}{\rm Tr}\left( i_k N^{(7)}+
i(2\pi\alpha^\prime)i_\Phi i_\Phi i_k N^{(9)}+\dots\right.\nonumber\\
&&+(2\pi\alpha^\prime)(i_k C^{(5)}+i(2\pi\alpha^\prime)
i_\Phi i_\Phi i_k N^{(7)}-\frac12 (2\pi\alpha^\prime)^2
(i_\Phi i_\Phi)^2 i_k N^{(9)})\wedge {\cal K}^{(2)}+\nonumber\\
&&\left. +(2\pi\alpha^\prime)(i_k B^{(6)}+i(2\pi\alpha^\prime)
i_\Phi i_\Phi i_k N^{(8)}-\frac12 (2\pi\alpha^\prime)^2
(i_\Phi i_\Phi)^2 i_k B^{(10)})\wedge {\cal K}^{(1)}
+\dots\right)\, .
\end{eqnarray}

\noindent The second term is associated to a configuration of N 
monopoles opening up into a so-called KK8-brane, exotic brane that
appears after double dimensionally reducing the
M9-brane \cite{HMO,EL}. (\ref{monoIIA}) generalizes the action for a
Kaluza-Klein monopole \cite{BEL} to the non-abelian 
case\footnote{Up to linear
terms in the Born-Infeld field strengths.} and shows that it also
contains couplings to higher dimensional spacetime fields, 
which can be interpreted in terms of
non-commutative brane configurations. 
The reduction of (\ref{Mmono2}) along a transverse direction 
different from the Taub-NUT direction produces the effective action
of a system of KK6-branes\footnote{See \cite{EL} for the abelian
case.} containing as well couplings to higher dimensional
spacetime fields, one of which can be interpreted as the KK6-branes
expanding into an NS9-brane.

The reduction of the effective action describing a system of M0-branes
gives rise to the effective action of non-abelian pp-waves in 
Type IIA. One finds:

\begin{equation}
\label{ppwaves}
S^{N waves}_{WZ}=\mu_0\int_R {\rm Tr}\left(k^{-2}k^{(1)}+i
(2\pi\alpha^\prime)i_\Phi i_\Phi (C^{(3)}-k^{-2}k^{(1)}\wedge
i_k C^{(3)})-\frac12 (2\pi\alpha^\prime)^2 (i_\Phi i_\Phi)^2
i_k B^{(6)}+\dots\right)\, .
\end{equation}

\noindent Here the second term indicates the existence of a configuration
corresponding to the waves expanding into a D2-brane transverse to
the direction in which they propagate, and the third term is
associated to pp-waves expanding into a wrapped NS5-brane.

In the next section we will see that the non-abelian brane
systems that we have derived in this section by reduction from M-theory
are connected
by T-duality to non-abelian Type IIB systems, predicted by the 
S-duality symmetry of the theory, consisting of strongly
coupled $p$-branes.  
Finally in section 8 we will give an interpretation of the terms that
couple to these actions through the Born-Infeld field strengths in
terms of topological solitons in brane-antibrane systems.

\section{Non-abelian Type IIB branes and S-duality}

In this section we discuss the interplay between S-duality and the
non-abelian effective actions for D$p$-branes derived in 
\cite{Myers} (see \cite{TvR} for a related discussion for D3-branes).
We will consider non-abelian systems of D$p$-branes,
with $p=1,3,5$, F1-branes and NS5-branes.

\subsection{Non-abelian D3-branes}

The worldvolume effective action describing a
single D3-brane is S-duality invariant. Although an S-duality
transformation maps the action into a different one, in which NS-NS
and RR 2-forms are interchanged and the abelian Born-Infeld field
describing open strings ending on the 3-brane is mapped into a
dual vector field associated to open D1-branes ending on it, 
one can see that 
this action is equivalent to the original one under a worldvolume
duality transformation that interchanges the two vector
fields. 
In the non-abelian case, however, the explicit worldvolume duality
transformation that connects weakly coupled and strongly coupled
D3-branes is not known.
As a consequence one seems to have independent 
worldvolume effective actions
to describe the system in the weak and strong coupling regimes.

Let us consider the following couplings in the Wess-Zumino action
of a U(N) system of D3-branes:

\begin{equation}
\label{D3weak}
S^{ND3}_{WZ}=\mu_3\int_{R^{3+1}}{\rm Tr}\left(C^{(4)}+
i(2\pi\alpha^\prime)i_\Phi i_\Phi C^{(6)}-\frac12 C^{(2)}\wedge
B^{(2)}+(2\pi\alpha^\prime)C^{(2)}\wedge {\cal F}+\dots\right)\, ,
\end{equation}

\noindent where we have chosen the basis in which $C^{(4)}$ is
S-duality invariant: $C^{(4)}\rightarrow C^{(4)}-\frac12 C^{(2)}
\wedge B^{(2)}$.
These terms are mapped under S-duality into:

\begin{equation}
\label{D3strong}
S^{ND3}_{WZ}=\mu_3\int_{R^{3+1}}{\rm Tr}\left(C^{(4)}+
i(2\pi\alpha^\prime)i_\Phi i_\Phi B^{(6)}+\frac12 C^{(2)}
\wedge B^{(2)}-(2\pi\alpha^\prime)B^{(2)}\wedge {\tilde {\cal F}}
+\dots\right)\, ,
\end{equation}

\noindent where ${\tilde {\cal F}}=2\partial {\tilde A}+
i[{\tilde A},{\tilde A}]+\frac{1}{2\pi\alpha^\prime} C^{(2)}$.

Let us recall how the worldvolume duality transformation
works in the abelian case (see for instance \cite{Tseytlin2}).
It proceeds in two steps. First one substitutes 
$dA+\frac{1}{2\pi\alpha^\prime}B^{(2)}$ by 
a gauge invariant field strength ${\cal F}$,
and adds a Lagrange multiplier term:

\begin{equation}
\label{multipli}
\mu_3 (2\pi\alpha^\prime)^2
\int_{R^{3+1}}({\cal F}-\frac{1}{2\pi\alpha^\prime}
B^{(2)})\wedge d{\tilde A}
\end{equation}

\noindent which imposes the constraint that 
${\cal F}-\frac{1}{2\pi\alpha^\prime}B^{(2)}$ 
derives from
a vector potential upon integration over
${\tilde A}$. This way one recovers the original action. On
the other hand the dual action is obtained through the equation
of motion for ${\cal F}$, which is given by a non-linear expression 
in terms of ${\tilde {\cal F}}$ and the spacetime fields 
due to the contribution of the Born-Infeld Lagrangian.
One readily sees however from (\ref{multipli}) that a term: 
$\int_{R^{3+1}}B^{(2)}\wedge d{\tilde A}$ will
couple in the dual action. Therefore the worldvolume duality
transformation substitutes the term 
$\int_{R^{3+1}}C^{(2)}\wedge {\cal F}$ in the worldvolume
effective action of the D3-brane by its S-dual:
$\int_{R^{3+1}}B^{(2)}\wedge {\tilde {\cal F}}$. 
The presence of the couplings $i_\Phi i_\Phi C^{(6)}$ and
$i_\Phi i_\Phi B^{(6)}$ in the non-abelian case makes clear
that the embedding scalars should transform as well, and very
non-trivially, under worldvolume duality. 
This is in agreement with the observations in \cite{TvR},
where the duality transformation properties of certain operators
containing the embedding scalars were derived.
However the explicit worldvolume duality transformation that connects
the two actions is not known.
Therefore, perturbative processes governed by open strings should be
described by (\ref{D3weak}) and non-perturbative ones, governed by
open D-strings, should be described by the strongly coupled action
(\ref{D3strong}). The second term in (\ref{D3weak}) indicates the
existence of a configuration corresponding to N D3-branes expanding
into a D5-brane, which represents the Higgs vacuum of \cite{PS}.
The confining vacuum is in turn represented by D3-branes polarized
into an NS5-brane \cite{PS}, which is described by the second term
in (\ref{D3strong}). Note that $B^{(6)}$ does not couple in the
weakly coupled action (\ref{D3weak}), and therefore the configuration
corresponding to the N D3-branes expanding into an NS5-brane cannot
be described at weak coupling. Consistently with this picture the
N D3-branes can also be represented at weak coupling as a single D5-brane
with N units of magnetic flux, associated to the coupling:
$S^{D5}\sim\int_{R^{5+1}} C^{(4)}\wedge F$ in the D5-brane effective action,
or as a single NS5-brane with N units of D1 flux, associated to the
coupling in the NS5-brane effective action:
$S^{NS5}\sim \int_{R^{5+1}}C^{(4)}\wedge {\tilde F}$, at strong coupling.

Let us now discuss the behavior under T-duality
of the action (\ref{D3strong}) representing
a set of D3-branes at strong coupling.
T-duality along a transverse direction
gives rise to the following couplings:

\begin{equation}
\label{actionD4}
S^{ND4_w}_{WZ}=\mu_3\int_{R^{3+1}}{\rm Tr}\left(i_k C^{(5)}+
i(2\pi\alpha^\prime)i_\Phi i_\Phi i_k N^{(7)}+\dots\right)\, ,
\end{equation}

\noindent corresponding to the wrapped D4-branes that we obtained
in the previous section from the reduction of the non-abelian
M5-branes.
A basic difference with the same operation in
the weakly coupled action (\ref{D3weak}), giving rise to D4-branes,
is that (\ref{actionD4}) cannot be unwrapped. At the level of the
terms that we have included this is a consequence of the T-duality
transformation:

\begin{equation}
\label{BN}
B^{(6)}_{\mu_1\dots\mu_6}\rightarrow (i_k N^{(7)})_{\mu_1\dots\mu_6}
+\dots
\end{equation}

\noindent for the NS-NS 6-form\footnote{This T-duality rule was derived
in \cite{EJL} as a key ingredient in order to prove the connection
between the Type IIB NS5-brane and the Type IIA Kaluza-Klein 
monopole under T-duality.}.
Therefore, we encounter again the situation in which the
same brane, in this case a D4-brane, is described by different
effective actions at weak and strong couplings.
As we mentioned already in the previous section  
(\ref{actionD4}) can indicate the existence of
a configuration in which (wrapped) D4-branes open up into a 
monopole, which cannot be explained however through the usual
couplings present in the weakly coupled D4-brane effective action.

Let us now consider a T-duality transformation
along a direction in the worldvolume of the N D3-branes at
strong coupling. We find N D2-branes that cannot move in the
direction of the duality transformation.
The relative minus sign of the $C^{(2)}\wedge B^{(2)}$ term
in (\ref{D3weak}) and (\ref{D3strong}) plays a key role in this
derivation. While in (\ref{D3weak})
the second term in the T-duality rule for $C^{(4)}$:

\begin{equation}
C^{(4)}_{\mu\nu\rho z}\rightarrow C^{(3)}_{\mu\nu\rho}-\frac32 
C^{(3)}_{[\mu\nu z}\frac{g_{\rho]z}}{g_{zz}}\, ,
\end{equation}

\noindent where $z$ denotes the direction of the T-duality
transformation,
is cancelled with the T-dual of $C^{(2)}\wedge B^{(2)}$, 
so that only the $C^{(3)}$ coupling of the D2-brane effective action
remains, the relative
minus sign of this term in (\ref{D3strong}) gives an overall
coupling: 
$C^{(3)}_{abc}-3 C^{(3)}_{[abz}\frac{g_{c]z}}{g_{zz}}=
C^{(3)}D_aXD_bXD_cX$. This, together with
the T-duality tranformation rule of $B^{(6)}$ (see \cite{EJL}):

\begin{equation}
B^{(6)}_{\mu_1\dots\mu_5 z}\rightarrow B^{(6)}_{\mu_1\dots\mu_5 z}+
\dots
\end{equation}

\noindent gives:

\begin{equation}
S^{ND2_t}_{WZ}=\mu_2\int_{R^{2+1}}{\rm Tr}\left(C^{(3)}DXDXDX+
i(2\pi\alpha^\prime)i_\Phi i_\Phi i_k B^{(6)}+\dots\right)\, ,
\end{equation}

\noindent which is the expression (\ref{actiontD2}) that we
derived from M-theory representing transverse D2-branes.
Again, this strongly coupled description of the D2-branes indicates
the existence of configurations that cannot be explained
by looking at the weakly coupled action. 
This is the case for the situation in which N D2-branes
expand into an NS5-brane, studied in \cite{Bena}.

The existence of these configurations is
predicted both from the analysis of the non-abelian systems
that we can construct in M-theory and the combined action of
S- and T-duality transformations for coinciding D3-branes.
We find new configurations corresponding to N D2-branes expanding 
into a wrapped NS5-brane \cite{Bena} and N D4-branes opening 
up into a Type IIA Kaluza-Klein monopole. 
The description in terms of the effective actions
that we have constructed is
valid at strong coupling, because the dynamics of both the
D2-branes and the wrapped D4-branes is governed by wrapped D2-branes
ending on them, as can be inferred both from the reduction from
M-theory and from the T-duality transformation
of the dual Born-Infeld field strength in (\ref{D3strong}).

\subsection{Non-abelian D1-branes}

Let us consider now a system of N coincident D1-branes. The 
corresponding effective action, according to Myers prescription,
contains the terms:

\begin{equation}
S^{N D1}_{WZ}=\mu_1\int_{R^{1+1}}{\rm Tr}
\left(C^{(2)}+i(2\pi\alpha^\prime)i_\Phi i_\Phi C^{(4)}
-\frac12 (2\pi\alpha^\prime)^2 (i_\Phi i_\Phi)^2 C^{(6)}
+\dots\right)\, .
\end{equation}

\noindent An S-duality transformation gives rise to the effective
action describing a system of coincident fundamental strings:

\begin{equation}
S^{N F1}_{WZ}=\mu_1\int_{R^{1+1}}{\rm Tr}
\left(B^{(2)}+i(2\pi\alpha^\prime)i_\Phi i_\Phi C^{(4)}
-\frac12 (2\pi\alpha^\prime)^2 (i_\Phi i_\Phi)^2 B^{(6)}
+\dots\right)\, .
\end{equation}

\noindent Here the second term indicates the existence of a configuration
corresponding to the N F1's opening up into a D3-brane,
configuration that has been studied in \cite{CMT}, and in \cite{Hasi}
from the point of view of the abelian D3-brane theory.

A T-duality transformation along a transverse direction to the N F1's
gives rise to the following action in Type IIA:

\begin{equation}
S^{N F1}_{WZ}=\mu_1\int_{R^{1+1}}{\rm Tr}
\left(B^{(2)}DXDX+i(2\pi\alpha^\prime)i_\Phi i_\Phi 
i_k C^{(5)}-\frac12 (2\pi\alpha^\prime)^2 (i_\Phi i_\Phi)^2 i_k N^{(7)}
+\dots\right)\, ,
\end{equation}

\noindent where the first term arises from the T-duality transformation
of $B^{(2)}$: $B^{(2)}_{ab}\rightarrow 
B^{(2)}_{ab}-2 B^{(2)}_{[a z}\frac{g_{b]z}}{g_{zz}}=
B^{(2)}D_aXD_bX$ and $B^{(6)}\rightarrow 
i_k N^{(7)}$ as in (\ref{BN}).
Note that these are
the same couplings present in (\ref{actiontF1}),
which was derived by reduction from M-theory. 
This action describes a delocalized fundamental string.
Again, in the abelian case a worldvolume duality transformation would
map this action into the action of an ordinary fundamental string.

Finally, T-duality along a worldvolume direction gives rise to N
pp-waves in Type IIA. We find the same couplings that we obtained in
the reduction from M-theory:

\begin{equation}
S^{N waves}_{WZ}=\mu_0\int_R {\rm Tr}\left(k^{-2}k^{(1)}+i
(2\pi\alpha^\prime)i_\Phi i_\Phi (C^{(3)}-k^{-2}k^{(1)}\wedge
i_k C^{(3)})-\frac12 (2\pi\alpha^\prime)^2 (i_\Phi i_\Phi)^2
i_k B^{(6)}+\dots\right)\, .
\end{equation}

\subsection{Non-abelian D5-branes}

Let us now perform the same analysis of the previous subsection 
but with a non-abelian system of D5-branes. The Wess-Zumino
action contains the terms:

\begin{eqnarray}
S^{N D5}_{WZ}&=&\mu_5\int_{R^{5+1}}{\rm Tr}\left(C^{(6)}+
i(2\pi\alpha^\prime)i_\Phi i_\Phi C^{(8)}+\dots\right.\nonumber\\
&&\left. +(2\pi\alpha^\prime)(C^{(4)}+i(2\pi\alpha^\prime)C^{(6)}+\dots)
\wedge {\cal F}+\dots\right)\, ,
\end{eqnarray}

\noindent from where S-duality gives:

\begin{eqnarray}
\label{NS5}
S^{N NS5}_{WZ}&=&\mu_5\int_{R^{5+1}}{\rm Tr}\left(B^{(6)}+
i(2\pi\alpha^\prime)i_\Phi i_\Phi 
{\tilde C}^{(8)}+\dots\right.\nonumber\\
&&\left. +(2\pi\alpha^\prime)(C^{(4)}+i(2\pi\alpha^\prime)B^{(6)}+
\dots)\wedge {\tilde {\cal F}}+\dots\right)\, .
\end{eqnarray}

\noindent Here ${\tilde C}^{(8)}$ is the field to which the 
D7-brane couples minimally at strong coupling (see \cite{EL}).
Therefore the second term represents the N NS5-branes opening
up into a D7-brane. 
Making now a T-duality transformation along a transverse direction,
we find:

\begin{equation}
S^{N KK}_{WZ}=\mu_5\int_{R^{5+1}}{\rm Tr}\left(i_k N^{(7)}+
i(2\pi\alpha^\prime)(i_\Phi i_\Phi i_k N^{(9)})+\dots\right)\, ,
\end{equation}

\noindent where we have used (\ref{BN}) and $i_k N^{(9)}$ 
arises from the dualization of ${\tilde C}^{(8)}$ (see \cite{EL}):

\begin{equation}
{\tilde C}^{(8)}_{\mu_1\dots\mu_8}\rightarrow 
(i_k N^{(9)})_{\mu_1\dots\mu_8}+\dots\, .
\end{equation}

\noindent This action describes N coincident Kaluza-Klein monopoles,
and we encountered it already in our reduction from M-theory.

T-dualizing (\ref{NS5}) along a worldvolume direction we find:

\begin{equation}
S^{N (NS5)_w}_{WZ}=\mu_4\int_{R^{4+1}}{\rm Tr}\left(i_k B^{(6)}+
i(2\pi\alpha^\prime)i_\Phi i_\Phi i_k N^{(8)}+\dots\right)\, ,
\end{equation}

\noindent where $i_k N^{(8)}$ arises from the dualization of 
${\tilde C}^{(8)}$ (see (3.3) in \cite{EL}):

\begin{equation}
{\tilde C}^{(8)}_{\mu_1\dots\mu_7 z}\rightarrow 
(i_k N^{(8)})_{\mu_1\dots\mu_7}+\dots\, .
\end{equation}

\noindent The KK6-brane of Type IIA couples
minimally to this field.
Therefore $i_\Phi i_\Phi i_k N^{(8)}$ indicates the existence
of a configuration of
N wrapped NS5-branes opening up into a KK6-brane, as we discussed in
the previous section.

\section{BPS solitons from non-abelian brane antibrane configurations}

The analysis of the couplings in the Wess-Zumino action 
describing a non-abelian 
brane-antibrane system reveals the emergence of new solitonic solutions
after the tachyonic mode of the open strings stretched between branes and
antibranes condenses\footnote{See \cite{reviews} for reviews on the role
of tachyonic excitations in unstable brane systems.}.  
This study reveals that typically non-commutative configurations can
also be described as topological configurations in non-abelian
brane-antibrane systems.

To see how these solutions arise let us start by recalling the abelian case. 

The Wess-Zumino part of the
effective action corresponding to a D$p$ brane-antibrane system 
contains couplings to the two U(1) fields of the brane and the 
antibrane as
well as to the complex tachyon field \cite{KW}. It includes in 
particular the very simple term:

\begin{equation}
\int_{R^{p+1}}C^{(p-1)}\wedge (dA^{(1)}-dA^{(1)\prime})
\end{equation}

\noindent where $dA^{(1)}$ and $dA^{(1)\prime}$ denote the 
Born-Infeld field strengths of the brane and the 
antibrane\footnote{Since we will be dealing with worldvolume
forms of different degree we indicate it explicitly for each
field.}. 
It is by now well understood that the tachyon condenses through
a Higgs-like mechanism in which the relative
U(1) field on the brane-antibrane system gets a mass 
and there is a localized magnetic
flux on $R^2$ that acts as a
source for the RR $(p-1)$-form field, 
as can be deduced from
the coupling above. This signals the emergence of a D$(p-2)$-brane
as the associated topological defect \cite{Sen1}, 
qualitative conclusion that is supported 
by the CFT analysis of the system \cite{Sen2,MS}.

The analysis of the
couplings in the Wess-Zumino action   
provides a hint on the possible solitonic objects
that can emerge after tachyonic condensation 
in situations in which string perturbation theory 
cannot be applied.
One can explain for instance the emergence of the
fundamental string as a solitonic solution in D$p$, anti-D$p$
systems \cite{Yi,HL1}, configuration that should exist as a
consequence of various duality arguments. A duality transformation
in the $p+1$ dimensional worldvolume of the D$p$, anti-D$p$ system
maps the Born-Infeld vector into a $(p-2)$-form. Therefore, in
this dual description the tachyonic field would be associated to
a D$(p-2)$-brane stretched between the brane and the antibrane, and
it would be charged under the relative $(p-2)$-form.
The fundamental string arises after the condensation of this tachyon
field, as can be seen 
qualitatively from the
analysis of the WZ term of the dualized brane-antibrane
effective action \cite{Yi,HL1}. In this action one finds a 
coupling (see \cite{HL1}):

\begin{equation}
\int_{R^{p+1}}B^{(2)}\wedge (dA^{(p-2)}-dA^{(p-2)\prime})\, .
\end{equation}

\noindent This indicates 
that the fundamental string will arise as a topological soliton 
when the dual tachyon condenses through a Higgs-like mechanism,
in which there is a localized magnetic flux in the transverse
$R^{p-1}$.

The dual Higgs mechanism explains the decoupling of the
overall $U(1)$ gauge group of the system at strong 
coupling\footnote{Due to the opposite orientation of brane and
antibrane one has: $dA^{(p-2)}-dA^{(p-2)\prime}=\,^*(dA^{(1)}+
dA^{(1)\prime})$.}. 
At weak coupling it is interpreted in terms of 
confinement of the overall $U(1)$,
with the fundamental string emerging as the confined electric
flux string at the end of the annihilation process \cite{BHY}.
This mechanism has been used to explain the fate of the unbroken
$U(1)$ on the worldvolume of the annihilating system \cite{Yi}.

Let us now consider N coincident D$p$ brane-antibrane pairs.
The Wess-Zumino action describing this system has been constructed
in \cite{KW}, though this reference does not include the 
non-commutative couplings
to the embedding scalars introduced by Myers. Our aim in this section
is to discuss the interpretation of precisely these terms for the
creation of branes as solitonic solutions. For our purposes we will just
need to consider the sum of the
Wess-Zumino terms for branes and antibranes, i.e. we will ignore the
contribution from the tachyon field.
 
Let us consider for instance the case of N coincident (D4, anti-D4)
brane pairs. The WZ effective
action up to linear terms in the non-abelian
Born-Infeld field strength includes:

\begin{eqnarray}
\label{actionD4antiD4}
&&S_{{\rm WZ}}^{{\rm (D4,\bar{D4})}}=\mu_4 \int_{R^{4+1}}
(2\pi\alpha^\prime){\rm Tr}\left(\left[C^{(3)}+
i(2\pi\alpha^\prime)i_\Phi i_\Phi
C^{(5)}-\frac12 (2\pi\alpha^\prime)^2 (i_\Phi i_\Phi)^2 
C^{(7)}\right]\wedge\right.\nonumber\\
&&\left.\wedge (F^{(2)}-F^{(2)\prime})\right)\, ,
\end{eqnarray}

\noindent where $F^{(2)}$, $F^{(2)\prime}$ are the Born-Infeld
field strengths for branes and antibranes.
Similar non-abelian configurations to the ones discussed in the
previous sections
are possible when one considers 
non-abelian (D$p$, anti-D$p$) systems. In this case the N pairs can 
open up into
a (D6, anti-D6) system or into a (D8, anti-D8). 
In turn, annihilation of these
brane configurations will produce solitonic solutions which have as
well an interpretation in terms of expanding branes. 
Qualitatively one can see the emergence of these configurations
from the analysis of the couplings in the WZ action describing the
N (D4, anti-D4) pairs of branes. The first term in (\ref{actionD4antiD4})
describes the usual realization of the D2-brane as a vortex solution
in a (D4, anti-D4) system, in this case N D2-branes from N (D4, anti-D4)
pairs\footnote{See the discussion below (\ref{actionD4antiD4cuad}).}.
The next term: 

\begin{equation}
\int_{R^{4+1}}{\rm Tr}\left[i_\Phi i_\Phi C^{(5)}\wedge (F^{(2)}-
F^{(2)\prime})\right]
\end{equation}

\noindent signals the emergence of a configuration representing the
N D2-branes opening up into a D4-brane, which would act as a source
for the RR 5-form. This configuration arises naturally from the 
annihilation via 
a Higgs-like mechanism of 
N (D4, anti-D4) pairs opening up into a (D6, anti-D6) system.
Similarly, the next term in (\ref{actionD4antiD4}) indicates the
existence of  
a non-commutative configuration of N D2-branes expanding into a
D6-brane, by-product of the annihilation of N (D4, anti-D4) pairs
expanding into a (D8, anti-D8) system. 

By analogy with the abelian case
one would expect that non-commutative configurations involving 
fundamental strings would arise after condensation of the
dual tachyon, charged with respect to the worldvolume dual of the
Born-Infeld field. However, the general mechanism for the dualization
of the non-abelian vector field is not known,
and not even a qualitative description
of non-abelian configurations of fundamental strings can be made
in these terms. We will see however in the next section that certain
configurations can be described from
non-abelian ($p$, ${\bar p}$) systems in M-theory.

Going back to (\ref{actionD4antiD4}),
including higher order terms in the Born-Infeld field strengths 
gives rise to new solitonic configurations. Bearing in 
mind that the topologically non-trivial character of the soliton 
can be carried by just one of the two field strengths, say $F^{(2)}$
(see for instance \cite{Sen2}),
the next contribution can be read from the action associated to
N D4-branes, namely from:

\begin{eqnarray}
\label{actionD4antiD4cuad}
S_{{\rm WZ,quad}}^{\rm D4}&=&\mu_4 \int_{R^{4+1}}
\frac12 (2\pi\alpha^\prime)^2{\rm Tr}\left(\left[C^{(1)}+
i(2\pi\alpha^\prime)i_\Phi i_\Phi
C^{(3)}-\frac12 (2\pi\alpha^\prime)^2 (i_\Phi i_\Phi)^2 
C^{(5)}+\dots\right]\wedge \right.\nonumber\\
&&\left. \wedge F^{(2)}\wedge F^{(2)}\right)\, .
\end{eqnarray}

\noindent The first term describes the realization of a 
D0-brane
as an instanton-like configuration in the N (D4, anti-D4) system.
$\int_{R^4} {\rm Tr}F^{(2)}\wedge F^{(2)}$ gives an integer iff
the homotopy group $\Pi_3(U(N))=Z$, which happens generically
for $N>2$. In the particular case $N=2^{k-1}$, where $2k$ denotes 
the codimension of the topological defect (in this case $k=2$)
it is possible to give a representation of the tachyon vortex
configuration (the generator of $\Pi_{2k-1}(U(N))$) such that all
higher and lower dimensional charges vanish \cite{Witten}. 
In this case
a given D$p$-brane can be realized as a bound state of $N$ D$(p+2k)$ 
brane-antibrane pairs stepwise, i.e. through a hierarchy of 
embeddings onto higher dimensional brane-antibrane systems.
Generically, for N D$(p+2k)$ brane-antibrane pairs this
representation of the tachyon should give rise to N/$2^{k-1}$
D$p$-branes as instanton-like configurations. Therefore, in
our case the N (D4, anti-D4) pairs would give rise to N/2
D0-branes\footnote{For this to make sense we need an even N.}.
The non-commutative configuration of N/2 D0-branes 
expanding into a D2-brane would arise as an instanton-like solution
in the worldvolume of N (D4, anti-D4) pairs expanding into
2 (D6, anti-D6) pairs, as implied by the second term in
(\ref{actionD4antiD4cuad}). Similarly, the third term in this
expression would describe the N/2 D0-branes expanding into a D4-brane
as a bound state of N (D4, anti-D4) pairs expanding into
2 (D8, anti-D8) pairs.

This discussion can be generalized to arbitrary N D$p$
brane-antibrane systems. One finds that a non-commutative
configuration of N/$2^{k-1}$ D$(p-2k)$-branes expanding into
a D$(p-2k+2r)$-brane\footnote{With N a multiple of $2^{k-1}$.}
would be realized as a bound state of
N (D$p$, anti-D$p$) branes opening up into $2^{k-1}$
(D$(p+2r)$, anti-D$(p+2r)$) branes.

A similar analysis can be performed in the Type IIB theory.
In this case it is easy to include as well non-abelian systems of 
NS-NS branes and antibranes. For instance, one easily sees from
(\ref{NS5}) that a non-abelian system of (NS5, anti-NS5) can
support N D3-branes expanding into an NS5-brane as a topological
configuration, arising after the condensation of the tachyon associated
to open D1-branes stretched between the NS5 and the anti-NS5 branes.
One notices as well that a system of coincident fundamental strings
arises as a topological soliton from a non-abelian (D3, anti-D3) brane
configuration at strong coupling, 
with open D1-branes stretched between the branes and the antibranes.
Including higher order terms in (\ref{D3strong}) one finds that
the system can also support a non-abelian configuration associated
to the N fundamental strings opening up into a D3-brane.  
Finally, although we did not consider D7-branes, one can easily
see that coincident (D7, anti-D7) branes may give rise to N
NS5-branes expanding into a D7-brane at strong coupling, i.e. when
the tachyonic mode associated to open D1-branes stretched between
branes and antibranes condenses.

\section{M-theory interpretation}

For a single brane-antibrane system one possible way of inferring 
the emergence of the fundamental string as a topological soliton
comes from M-theory \cite{Yi,HL1}. Let us consider for instance
a (D4, anti-D4) pair in Type IIA. This system corresponds in M-theory
to a coincident (M5, anti-M5) pair. 
The tachyonic mode in the
open string stretched between the D4 and the anti-D4 branes occurs
in M-theory in the form of a tachyonic mode in an M2-brane stretched 
between the M5 and the anti-M5 branes.
The condensation of this tachyon through the previously discussed
Higgs mechanism
gives rise to an
M2-brane as the resulting topological defect, as can be inferred both
from the duality with Type IIA and from the term:

\begin{equation}
\int_{R^{5+1}}{\hat C}\wedge d{\hat a}^{(2)}
\end{equation}

\noindent in the worldvolume effective action of the (M5, anti-M5)
system \cite{Yi}. Here ${\hat a}^{(2)}$ is the
six dimensional worldvolume antisymmetric tensor, which for the
brane antibrane pair is not constrained by self or anti-self
duality (see \cite{Yi}).
As Yi pointed out the reduction along a
direction in the worldvolume of the (M5, anti-M5) pair transverse 
to the
stretched M2-brane gives rise to a situation in which the tachyonic
mode of an open D2-brane stretched between a D4 and an anti-D4 brane 
condenses to give rise to a fundamental string as the topological
defect. This is predicted by the coupling above, where 
${\hat a}^{(2)}$ reduces to a 2-form worldvolume field coupling 
to an open D2-brane in Type IIA, and 
$B^{(2)}_{\mu\nu}={\hat C}_{\mu\nu 11}$ remains as the
resulting field.
A systematic
study of the possible brane-antibrane configurations in M-theory
and Type II theories can be found in \cite{HL1}, where it is seen 
that Kaluza-Klein reduction of the different possible M-theory 
configurations predicts that generically
fundamental strings arise as topological defects in (D$p$, anti-D$p$)
systems after the tachyonic mode of a D$(p-2)$-brane stretched
between the pair condenses, and that
the NS5-brane, the wave, the Kaluza-Klein
monopole and the so-called exotic branes may also arise 
as topological solitons from various types of configurations.

We can now consider non-abelian brane-antibrane systems and
try to provide a similar description.
Let us start by considering N coincident 
(D4, anti-D4) pairs. The situation in which the 
N pairs expand into a (D6, anti-D6) pair,
which supports as solitonic configuration N D2-branes opening up into
a single D4-brane, can be described in M-theory as N pairs of
wrapped (M5, anti-M5) branes opening up into a Kaluza-Klein
monopole-antimonopole pair. 
The Wess-Zumino action describing N (M5, anti-M5) pairs
contains the couplings:

\begin{equation}
\label{M5antiM5}
S^{N({M5}_w, {\bar M5}_w)}_{\rm WZ}=\mu_4\int_{R^{4+1}}{\rm Tr}
\left(l_p^2\left[{\hat C}
D{\hat X}D{\hat X}D{\hat X}+il_p^2 i_{\hat \Phi} i_{\hat \Phi} 
i_{\hat k} {\hat {\tilde C}}+
\dots\right]\wedge ({\hat F}^{(2)}-{\hat F}^{(2)\prime})\right)\, ,
\end{equation}

\noindent as implied from (\ref{actionNM5}).
These couplings describe a solitonic
configuration representing N M2-branes opening up into a wrapped
M5-brane, with the M2-branes delocalized in the direction on which
the M5-brane is wrapped.
This configuration occurs when the tachyonic mode of the
wrapped M2-branes stretching between the M5 and the anti-M5 branes
condenses, and is consistent with the fact that a wrapped M5-brane
can arise as a soliton after tachyonic condensation of a single
(M6, anti-M6) pair, into which the system of N 
($M5_w$, anti-$M5_w$) branes
expands. It is again consistent that the possible M2-branes that
can end on both the wrapped M5-branes and the M6-branes are wrapped
M2-branes. Had the M5-branes been unwrapped we would have found an
inconsistency in this description derived from the fact that only
unwrapped M2-branes could have ended on them.
 
Let us now discuss the case in which N (D2, anti-D2) branes
in Type IIA annihilate to give rise to N D0-branes. This system
supports a solitonic configuration corresponding to the N D0
branes opening up into a single D2-brane, which is described by the
coupling:

\begin{equation}
\label{D2antiD2}
\int_{R^{2+1}}{\rm Tr}\left(i_\Phi i_\Phi C^{(3)}\wedge 
(F^{(2)}-F^{(2)\prime})\right)\, ,
\end{equation}

\noindent in the Wess-Zumino action of the (D2, anti-D2) branes
(see (\ref{actionND2})).
The M-theory description of a single (D2, anti-D2) pair
consists on an (M2, anti-M2) system in which the tachyonic
mode of a wrapped M2-brane stretched between the brane and the
antibrane condenses, giving rise to an M-wave as the topological
defect, moving in the
same direction on which the stretched M2-brane is wrapped \cite{HL1}.
This is described by the coupling:

\begin{equation}
\int_{R^{2+1}} {\hat k}^{-2}{\hat k}^{(1)}\wedge 
({\hat F}^{(2)}-{\hat F}^{(2)\prime})
\end{equation}

\noindent in the (M2, anti-M2) effective action.
In the non-abelian case the Wess-Zumino action of the 
N (M2, anti-M2) system contains the couplings:

\begin{equation}
S^{N (M2_t, \bar{M2}_t)}_{WZ}=\mu_2\int_{R^{2+1}}{\rm Tr}
\left(l_p^2\left[{\hat k}^{-2}{\hat k}^{(1)}+il_p^2 i_{\hat \Phi}
i_{\hat \Phi}({\hat C}-{\hat k}^{-2}{\hat k}^{(1)}
\wedge i_{\hat k}{\hat C})+\dots\right]\wedge
({\hat F}^{(2)}-{\hat F}^{(2)\prime})+\dots\right)\, .
\end{equation}

\noindent The second term indicates that a non-trivial
localized magnetic flux on $R^2$ gives rise to
a configuration corresponding
to M0-branes opening up into a transverse M2-brane as
a topological solution.
As we discussed before, this is precisely the kind of M2-brane that
can occur as a topological
defect in an (M5, anti-M5) system in which the M5-branes are wrapped
on the special direction transverse to the M2-brane. 
This is then consistent with the picture in which the N (M2, anti-M2)
pairs would open up into a single (M5, anti-M5) pair giving 
N M0-branes expanding into an
M2-brane as the resulting topological defect.

The analysis of the effective action describing N coincident 
(KK, anti-KK) pairs reveals the existence of a non-commutative
configuration corresponding to N M5-branes expanding into a
monopole as a topological defect, with the M5-branes wrapped in
the Taub-NUT direction of the monopole. This is consistent with
the situation in which the N (KK, anti-KK) pairs open up into
an (M9, anti-M9) pair, that supports a Kaluza-Klein monopole as
a topological defect (see \cite{HL1}).
Reducing this configuration along the
Killing direction one obtains a configuration of N D4-branes
opening up into a D6-brane, which we discussed in the previous
section.

\section{Other brane-antibrane configurations in Type IIA}

The M-theory configurations that we have considered 
in the previous section were constructed in such a way that they 
reproduced non-abelian (D$p$, anti-D$p$) systems when
reduced along their isometric direction. We are now going to see
that it is possible
to obtain other interesting configurations in Type IIA after reduction
along a different direction. The solitonic configurations that we
find in this section are connected via T-duality with the strongly
coupled configurations of Type IIB solitonic branes that we considered
in section 6.

Let us start by considering a system of coincident
(M5, anti-M5) branes. Reducing along a transverse direction gives rise
to coincident (NS5, anti-NS5) pairs, wrapped in some special direction.
The leading term of the corresponding effective action is given by
the second line in (\ref{actionNS5}), with ${\cal H}^{(2)}$ replaced 
by the relative field strength. The second term shows
that a configuration corresponding to
N (localized) D2-branes expanding into a (wrapped) NS5-brane can 
arise as a topological solution. This is consistent with
the situation in which the N (NS5, anti-NS5) pairs expand into a
pair of (KK6, anti-KK6) branes, since this single pair supports an
NS5-brane as a topological solution, with the brane wrapped in the
special Killing direction of the KK6-brane (see \cite{HL1}).
On the other hand, double dimensional reduction of the N (M5, anti-M5)
system gives rise to N (D4, anti-D4) pairs. The leading terms of the 
worldvolume effective action are given by the second and third lines
in (\ref{actionD4w}). The coupling to the two different field strengths
${\cal K}^{(2)}$ and ${\cal K}^{(1)}$ shows that the system can
support two different types of solitonic configurations depending on
which type of stretched brane has its tachyonic mode condensing.
If the tachyonic mode
of stretched wrapped D2-branes condenses one ends up with a
configuration corresponding to 
N fundamental strings expanding into
a D4-brane. This is described by the second coupling in the third line of
(\ref{actionD4w}).
On the other hand if the stretched branes are fundamental strings
then N D2-branes expanding into an NS5-brane arise as the topological
defect. This is described by the second coupling in the second line of
(\ref{actionD4w}).
In both
cases the N branes are transverse to the special direction in which
the expanded brane is wrapped. As before, these configurations are
consistent with the situation in which N (D4, anti-D4) pairs
expand into a (KK, anti-KK) pair, since the latter
can support D4 and NS5 branes, wrapped in the Taub-NUT direction,
as topological solutions \cite{HL1}.

A similar analysis starting with a delocalized (M2, anti-M2) 
non-abelian system 
gives rise to the following configurations. Direct dimensional reduction
gives N coincident delocalized (D2, anti-D2) branes, which can support N
pp-waves expanding into a transverse D2-brane as a topological solution.
The coupling 
$\int_{R^{2+1}}{\rm Tr}{\hat k}^{-2}{\hat k}^{(1)}
\wedge (db^{(1)}-db^{(1)\prime})$, present in the worldvolume effective
action of the system (see (\ref{actiontD2})),
shows that N pp-waves may arise as a topological defect for
non-vanishing magnetic flux. Recall that this magnetic flux
is associated to open, wrapped, D2-branes stretched between
branes and antibranes. For a localized (D2, anti-D2) system one can
only see the emergence of a pp-wave as a solitonic solution
for a single pair. The reason is that one needs to perform a
worldvolume duality transformation in order to identify the
coupling responsible for this process. Recalling the results in
\cite{HL1}, one first needs to select one of the transverse
directions to the D2-brane system, and write the coupling to the
3-form RR-potential as:
$\int_{R^{2+1}}C^{(3)}=\int_{R^{2+1}}(C^{(3)}+i_k C^{(3)}\wedge dy)$.
Then $y$ is dualized into a 
worldvolume vector field by adding a Lagrange multiplier term:
$\int_{R^{2+1}}dy\wedge db^{(1)}$, and the final dual action contains
the term responsible for the creation of the pp-wave:
$\int_{R^{2+1}}k^{-2}k^{(1)}\wedge db^{(1)}$ (see \cite{HL1}).
For a non-abelian configuration of D2-branes one cannot however
see the emergence of this coupling, given that the explicit
worldvolume duality transformation cannot be made. One needs to
start with a configuration in which the direction of propagation
of the wave is already singled out, as is the case for
a configuration of delocalized D2-branes.
The emergence of this non-commutative configuration
is consistent with the situation in which the transverse D2-branes
open up into an NS5-brane, which can support a transverse D2-brane as a 
topological soliton, as we have discussed above. 

Reducing the (M2, anti-M2) system along a 
worldvolume direction one obtains N coincident (F1, anti-F1) branes
delocalized in one direction.
This system supports as well a topological defect corresponding to
N pp-waves opening up into a D2-brane, 
now arising after the tachyon of the 
stretched, wrapped, fundamental strings condenses. 
This can be seen from the second line in (\ref{actiontF1}).
This configuration is
consistent with the fact that the transverse F1-branes
can open up into a ($D4_w$, anti-$D4_w$) pair, and this system 
admits a delocalized D2-brane as a topological soliton, as we
have described above.

Finally, the reduction of a non-abelian system of coincident Kaluza-Klein
anti Kaluza-Klein monopoles gives, when reducing along a worldvolume
direction, a system of coincident Type IIA (KK, anti-KK) pairs, which
can support two types of solitonic configurations: N D4-branes
expanding into a monopole, and N NS5-branes expanding into a KK6-brane,
depending on whether the condensing tachyon is associated to open,
wrapped, D2-branes or F1-branes. 
The corresponding couplings can be read from (\ref{monoIIA}). 
This is consistent with the situation in which the Kaluza-Klein
monopoles expand into a (KK8, anti-KK8) pair, since this system can
support both a Kaluza-Klein monopole and a KK6-brane as solitonic
solutions (see \cite{HL1}).
The reduction along a transverse direction gives a system
of coincident (KK6, anti-KK6) branes, which supports as a topological
defect a configuration corresponding to N NS5-branes opening up
into a KK6-brane. This is consistent with the situation in which
N (KK6, anti-KK6) pairs expand into an (NS9, anti-NS9) pair, which can
support a KK6-brane as a topological defect \cite{HL1}.

\subsection*{Acknowledgements}

It is a pleasure to thank Laurent Houart for very useful discussions.

\end{document}